# A new Hysteretic Nonlinear Energy Sink (HNES)

## George C. Tsiatas[a], Aristotelis E. Charalampakis[b]

[a] Department of Mathematics, University of Patras, Rio, 26504, Greece.
e-mail: gtsiatas@math.upatras.gr, web page: http://www.math.upatras.gr/~gtsiatas/
[b] School of Civil Engineering, National Technical University of Athens, Athens, 15773, Greece.
e-mail: achar@mail.ntua.gr, web page: http://www.charalampakis.com/

**Abstract**

The behavior of a new Hysteretic Nonlinear Energy Sink (HNES) coupled to a linear primary oscillator is investigated in shock mitigation. Apart from a small mass and a nonlinear elastic spring of the Duffing oscillator, the HNES is also comprised of a purely hysteretic and a linear elastic spring of potentially negative stiffness, connected in parallel. The Bouc-Wen model is used to describe the force produced by both the purely hysteretic and linear elastic springs. Coupling the primary oscillator with the HNES three nonlinear equations of motion are derived, in terms of the two displacements and the dimensionless hysteretic variable, which are integrated numerically using the analog equation method. The performance of the HNES is examined by quantifying the percentage of the initially induced energy in the primary system that is passively transferred and dissipated by the HNES. Remarkable results are achieved for a wide range of initial input energies. The great performance of the HNES is mostly evidenced when the linear spring stiffness takes on negative values.

## 1. Introduction

Passive vibration suppression is an open research field which finds application to numerous scientific areas including acoustic, aeroelasticity, earthquake, mechanical and aerospace engineering. One of the oldest passive vibration control devices is the Tuned Mass Damper (TMD) consisting of a mass, a spring and a viscous damper which is attached to a primary vibrating system in order to suppress undesirable vibrations. The natural frequency of the TMD is usually tuned in resonance with the fundamental mode of the primary structure. Thus, a large amount of the structural vibrating energy is transferred to the TMD and then dissipated by damping. The TMD concept was proposed by Watts as early as 1883 [1] and patented by Frahm [2]. Even though TMDs are known for their effectiveness they possess certain drawbacks. First, environmental influences and other external parameters may alter the TMD properties, disturbing its tuning. Consequently, its performance can be significantly reduced [3]. Second, a large oscillating mass is generally required in order to achieve significant vibration reduction, rendering its construction and placement rather difficult.

In light of the above, new nonlinear strategies in vibration absorption have been introduced which overcome the disadvantages of linear TMDs. Among them, Nonlinear Energy Sink (NES) has gained tremendous attention by researchers who investigated the Targeted Energy Transfer (TET) through different designs of NESs [4]. In principle, the NES is a nonlinear attachment coupled to a linear system which irreversibly dissipates energy through nonlinear stiffness elements. NESs, unlike TMDs, do not need to be tuned to a particular frequency in order to dissipate energy effectively. Instead, they absorb energy at a wider range of frequencies and, therefore, are more robust than TMDs [5]. In the literature, seven types of NESs have been recorded so far. In Type I, II and III NES designs, an essential nonlinear cubic spring has been employed with linear (Type I and III) or nonlinear (Type II) damping [6]-[13]. In Type IV NES design, a rotating NES has been introduced coupled to a primary linear oscillator through an essentially nonlinear inertial nonlinearity [14]-[17], whereas Type V and VI designs are devoted to strongly nonlinear vibro-impact coupling [18]-[23]. The only NES which has incorporated negative stiffness elements was introduced by AL-Shudeifat [4] and has been realized through the geometric nonlinearity of the transverse linear springs. This NES design, known also as Type VII NES, proved that the negative stiffness elements considerably enhance the NES performance for passive energy pumping and local rapid energy dissipation.

In this work, the behavior of a new Hysteretic Nonlinear Energy Sink (HNES) coupled to a linear primary oscillator is investigated in shock mitigation. HNES can be considered as a modification of Type I NES [7], i.e., apart from a small mass and a nonlinear elastic spring of the Duffing oscillator, the HNES is also comprised of a purely hysteretic spring and a linear elastic spring of potentially negative stiffness, connected in parallel. The Bouc-Wen model is used to describe the force produced by the purely hysteretic and linear elastic springs. Coupling the primary oscillator with the HNES two nonlinear equations of motion are derived, as well as one evolution equation for the hysteretic parameter, which are integrated numerically using the analog equation method. The performance of the HNES is examined by quantifying the percentage of the initially induced energy in the primary system that is passively transferred and dissipated by the HNES. Remarkable results are achieved for a wide range of initial input energies. The great performance of the HNES is mostly evidenced when the linear spring constant takes on negative values.

## 2. Hysteretic element

The Bouc-Wen model, introduced by Bouc [24] and extended by Wen [25], is employed to describe the force produced by both the purely hysteretic and linear elastic springs. It is a very concise model that can be applied in hysteretic phenomena across diverse scientific fields, ranging from magnetism and elastoplasticity to sociology and medicine. Although developed independently, the Bouc-Wen model follows the endochronic theory pioneered by Valanis [26], which discards the notion of a yield surface to describe plasticity [27]. It captures the rate-independent hysteresis by means of internal (non-measurable) hysteretic variables which follow suitable differential equations.

For the case of a single-degree-of-freedom (SDoF) system, and following the notation in [28], [29], the hysteresis is expressed by

$$F^{BW}(t) = a\,k\,x(t) + (1-a)\,k\,D\,z(t), \qquad (1)$$

where, $k > 0$, $D > 0$, $x(t)$ is the time history of the input variable and $z(t)$ is a dimensionless hysteretic variable which is governed by the differential equation

$$\dot{z} = D^{-1}\left(A - \left(\beta\,\mathrm{sgn}(z\,\dot{x}) + \gamma\right)|z|^n\right)\dot{x}, \qquad (2)$$

in which $n > 0$ and $\mathrm{sgn}(\cdot)$ is the signum function. In the context of structural mechanics, $x$ is usually the displacement, $k$ the initial stiffness, $a$ the ratio of post to pre-yield stiffness, $D$ is the yield displacement and the dimensionless exponential parameter $n$ governs the abruptness of transition between pre- and post-yield response. The dimensionless parameters $A$, $\beta$, $\gamma$ control the shape and size of the hysteretic loop. For reasons of model consistency, the constraints $A = \beta + \gamma = 1$ are imposed which simplify the model without limiting its capabilities [29], leading to $z_{max} = \left(A/(\beta+\gamma)\right)^{1/n} = 1$.

Based on Eq. (1), the model can be visualized as two springs connected in parallel, i.e. a linear elastic and a purely hysteretic spring, with $F^{el}(t) = a\,k\,x(t)$ and $F^h(t) = (1-a)\,k\,D\,z(t)$, respectively. The combined response is shown in Fig. 1.

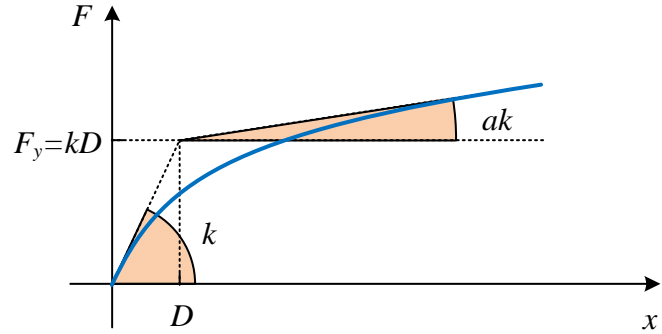

**Fig. 1.** Response of Bouc-Wen model under displacement-controlled monotonic loading.

It is known that the response of the Bouc-Wen model exhibits displacement drift and force relaxation when subjected to short unloading–reloading paths. Consequently, it locally violates Drucker's or Ilyushin's postulate of plasticity. This problem has been eliminated in a corrected model using an appropriate stiffening factor [29], [30], and has been implemented herein. The correction mostly affects the small energy experiments.

Regarding the dissipated energy of the Bouc-Wen model, this is expressed numerically by the area enclosed by hysteretic loops [29]. Formally, the proof is based on the 1st and 2nd laws of thermodynamics with the additional assumptions of (a) isothermal conditions and (b) same state of internal variables in the beginning and end of the cycle [32]. In Fig. 2, only the hysteretic spring is accounted for as the elastic spring does not dissipate energy (in general), nor does it store energy in a closed displacement loop.

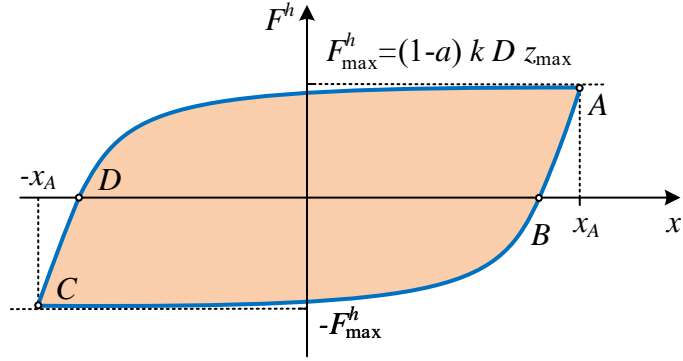

**Fig. 2.** Steady-state response of hysteretic spring under symmetric wave T-periodic excitation [29].

The dissipated energy between two arbitrary time instants is estimated numerically by the work done by the purely hysteretic spring, as

$$E_{DISS\_hys} \cong \int_{x_0}^{x} F^h dx = (1-a) D k \int_{t_0}^{t} (z \dot{x}) dt . \qquad (3)$$

The above equation is not exact in the sense that the hysteretic spring of the Bouc-Wen model actually stores some small recoverable energy. In case of a complete cycle, this energy is fully recovered in the unloading transitions A→B and C→D. Although exact calculation of the dissipated energy for an arbitrary transition is feasible on the basis of Gauss' hypergeometric functions [29], its implementation is cumbersome. Since the error is small and does not accumulate over time, Eq. (3) is used herein.

## 3. Negative stiffness element

The idea of employing negative stiffness springs, or 'anti-springs', for the absorption of oscillations can be traced in aeronautical engineering and the innovative paper by Molyneaux [33]. This idea was extended significantly by Platus [34]. Apart from pre-compressed springs, physical implementation of negative stiffness elements can be achieved by beams, slabs or shells in post-buckled arrangements, inverse pendulum systems, etc. Some interesting implementations of such non-linear isolation systems can be found in the works of Winterflood et al. [35], Virgin et al. [36], Liu et al. [37], Antoniadis et al. [38], and others.

Instead of introducing an additional negative stiffness element, a simplest approach is proposed herein which involves the already present linear elastic spring of the Bouc-Wen model. This spring can easily obtain negative stiffness for negative values of its parameter $a$, leading to a true softening behavior. The combined effect of all springs is shown in Fig. 3, which can be compared to Fig. 1. This approach allows for a seamless investigation of the oscillator's behavior for both positive and negative values of $a$.

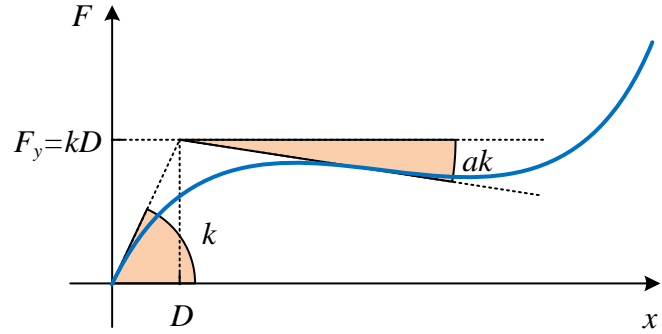

**Fig. 3.** Response of the proposed HNES under displacement-controlled monotonic loading ($a < 0$).

## 4. Dynamics of a SDoF primary system with HNES

The schematic representation of a SDoF linear primary system with HNES is shown in Fig. 4. The equations of motion for this system are:

$$m_1\ddot{u}_1 + c_1\dot{u}_1 + k_1 u_1 - c_2(\dot{u}_2 - \dot{u}_1) - F^{BW} - k_{nl2}(u_2 - u_1)^3 = 0, \quad (4)$$

$$m_2\ddot{u}_2 + c_2(\dot{u}_2 - \dot{u}_1) + F^{BW} + k_{nl2}(u_2 - u_1)^3 = 0, \quad (5)$$

where $m_1$, $u_1$, $k_1$, $c_1$ are the mass, the displacement, the stiffness, and the damping coefficient of the linear primary system, respectively; $m_2$, $u_2$, $c_2$, $k_{nl2}$ are the mass, the displacement, the damping coefficient, and the cubic nonlinear stiffness of the HNES attachment. The hysteretic restoring force $F^{BW}$ is derived according to Eq. (1) as

$$F^{BW} = ak_2 x(t) + (1-a)k_2 Dz(t), \quad (6)$$

where, $k_2 > 0$ is the initial stiffness of the whole Bouc-Wen model, $x \equiv u_2 - u_1$ is the input variable and the evolution of $z$ follows Eq. (2). In total, the equations of motion take the form

$$m_1\ddot{u}_1 + c_1\dot{u}_1 + k_1 u_1 - c_2(\dot{u}_2 - \dot{u}_1) - \left(ak_2(u_2 - u_1) + (1-a)k_2 Dz\right) - k_{nl2}(u_2 - u_1)^3 = 0, \quad (7)$$

$$m_2\ddot{u}_2 + c_2(\dot{u}_2 - \dot{u}_1) + \left(ak_2(u_2 - u_1) + (1-a)k_2 Dz\right) + k_{nl2}(u_2 - u_1)^3 = 0, \quad (8)$$

$$\dot{z} = D^{-1}(\dot{u}_2 - \dot{u}_1)\left(1 - |z|^n\left(\gamma\,\mathrm{sgn}((\dot{u}_2 - \dot{u}_1)z) + \beta\right)\right). \quad (9)$$

In this investigation, only direct impulsive forcing of the primary system is considered. Thus, the initial conditions of the problem are

$$\dot{u}_1(0) \neq 0, \text{ and } u_1(0) = \dot{u}_2(0) = u_2(0) = 0. \quad (10)$$

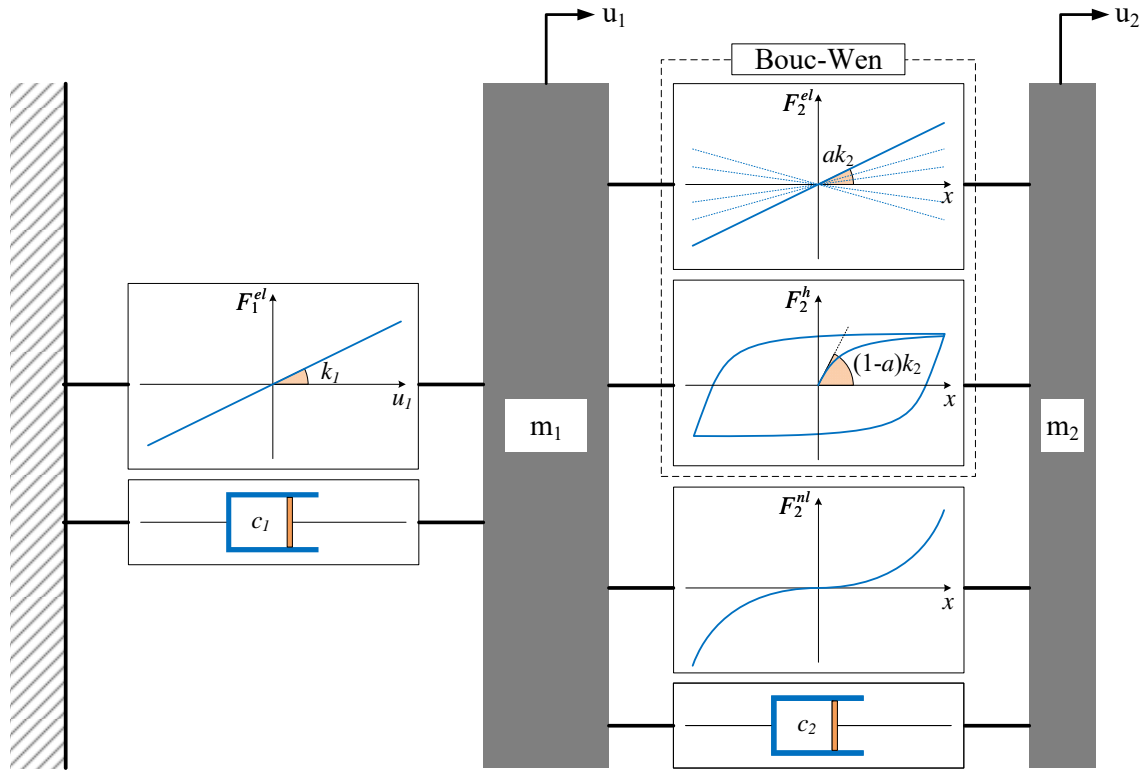

**Fig. 4.** A SDoF primary system with HNES.

## 5. State space form of the equations and their solution

The equations (7) - (9) can be recast into state space form as

$$\begin{cases} x_1 = u_1 \\ x_2 = \dot{u}_1 \\ x_3 = u_2 \\ x_4 = \dot{u}_2 \\ x_5 = z \end{cases}, \tag{11}$$

$$\begin{cases} \dot{x}_1 = x_2 \\ \dot{x}_2 = \dfrac{-k_1 x_1 - c_1 x_2 + c_2(x_4 - x_2) + \left(ak_2(x_3 - x_1) + (1-a)k_2 D x_5\right) + k_{nl2}(x_3 - x_1)^3}{m_1} \\ \dot{x}_3 = x_4 \\ \dot{x}_4 = \dfrac{-c_2(x_4 - x_2) - \left(ak_2(x_3 - x_1) + (1-a)k_2 D x_5\right) - k_{nl2}(x_3 - x_1)^3}{m_2} \\ \dot{x}_5 = D^{-1}(x_4 - x_2)\left[1 - |x_5|^n \left(\gamma\, sign\left((x_4 - x_2)x_5\right) + \beta\right)\right] \end{cases}, \tag{12}$$

together with the pertinent initial conditions

$$\begin{cases} \dot{x}_1 = v_0 \\ \dot{x}_2 = -\dfrac{(c_1 + c_2)v_0}{m_1} \\ \dot{x}_3 = 0 \\ \dot{x}_4 = \dfrac{c_2 v_0}{m_2} \\ \dot{x}_5 = \dfrac{-v_0}{D} \end{cases}, \tag{13}$$

where $v_0 = \dot{u}_1(0)$.

Eqs. (11) - (13) constitute a nonlinear initial value problem which is solved numerically using a new direct time integration method introduced by Katsikadelis [39]. The method is based on the principle of the analog equation, which converts the $N$ coupled equations into a set of $N$ single term uncoupled first order ordinary differential equations under fictitious sources. The solution is obtained from the integral representation of the solution of the substitute single term equations. Moreover, the method is simple to implement, it is self-starting, unconditionally stable, accurate, and it does not exhibit numerical damping [39].

## 6. Performance index

The performance of the HNES is examined by quantifying the percentage of the initially induced energy in the primary system $E_0 = m_1 v_0^2 / 2$ that is passively transferred and dissipated by the HNES. That is

$$\eta = 100 \frac{E_{DISS\_HNES}}{E_0}, \qquad (14)$$

where $E_{DISS\_HNES} = E_{DISS\_c2} + E_{DISS\_hys}$ is the total energy dissipated by the HNES, comprised of the viscous damper dissipated energy [40]

$$E_{DISS\_c2} = c_2 \int_0^t \left( \dot{u}_2(\tau) - \dot{u}_1(\tau) \right)^2 d\tau, \qquad (15)$$

and the hysteretic dissipated energy, based on Eq. (3),

$$E_{DISS\_hys} = (1-a) k_2 D \int_0^t z(\tau) \left( \dot{u}_2(\tau) - \dot{u}_1(\tau) \right) d\tau. \qquad (16)$$

Moreover, the instantaneous energy stored in the primary system can be written as

$$E_{INST\_PR} = \frac{1}{2} m_1 \dot{u}_1^2 + \frac{1}{2} k_1 u_1^2, \qquad (17)$$

whereas, the energy dissipated by the viscous damper of the primary system is expressed as

$$E_{DISS\_c1} = c_1 \int_0^t \dot{x}_1(\tau)^2 d\tau. \qquad (18)$$

Finally, the instantaneous energy stored in the HNES is given by

$$E_{INST\_HNES} = \frac{1}{2} m_2 \dot{u}_2^2 + \frac{1}{2} a k_2 (u_2 - u_1)^2 + \frac{1}{4} k_{nl2} (u_2 - u_1)^4. \qquad (19)$$

Evidently, the law of conservation of energy reads

$$E_0 = E_{INST\_PR} + E_{DISS\_c1} + E_{INST\_HNES} + E_{DISS\_HNES}. \qquad (20)$$

## 7. Numerical results and discussion

In this section, numerical results are presented which validate the effectiveness of the HNES for a wide range of initial input energies.

*7.1 Numerical validation - Duffing oscillator coupled to a NES*

In order to validate our numerical solution to the nonlinear initial value problem, a Duffing oscillator coupled to a Type I NES is first examined and the results are compared to those obtained by Viguié et al. [40]. In this case, the equations of motion are given by [40]

$$m_1 \ddot{u}_1 + c_1 \dot{u}_1 + k_1 u_1 + k_{nl1} u_1^3 - c_2 (\dot{u}_2 - \dot{u}_1) - k_{nl2} (u_2 - u_1)^3 = 0, \quad (21)$$

$$m_2 \ddot{u}_2 + c_2 (\dot{u}_2 - \dot{u}_1) + k_{nl2} (u_2 - u_1)^3 = 0, \quad (22)$$

and the employed data are: $m_1 = 1\,\text{kg}$, $m_2 = 0.05\,\text{kg}$, $k_1 = 1\,\text{N/m}$, $k_{nl1} = 9\,\text{N/m}^3$, $k_{nl2} = 1\,\text{N/m}^3$, and $c_1 = c_2 = 0.002\,\text{Ns/m}$. Fig. 5 depicts the Duffing oscillator and the NES responses for time series of 500s and 50s, the percentage of instantaneous energy stored in the NES, as well as the percentage of initial energy dissipated in the NES. It can be shown that the results practically coincide with those obtained by Viguié et al. [40].

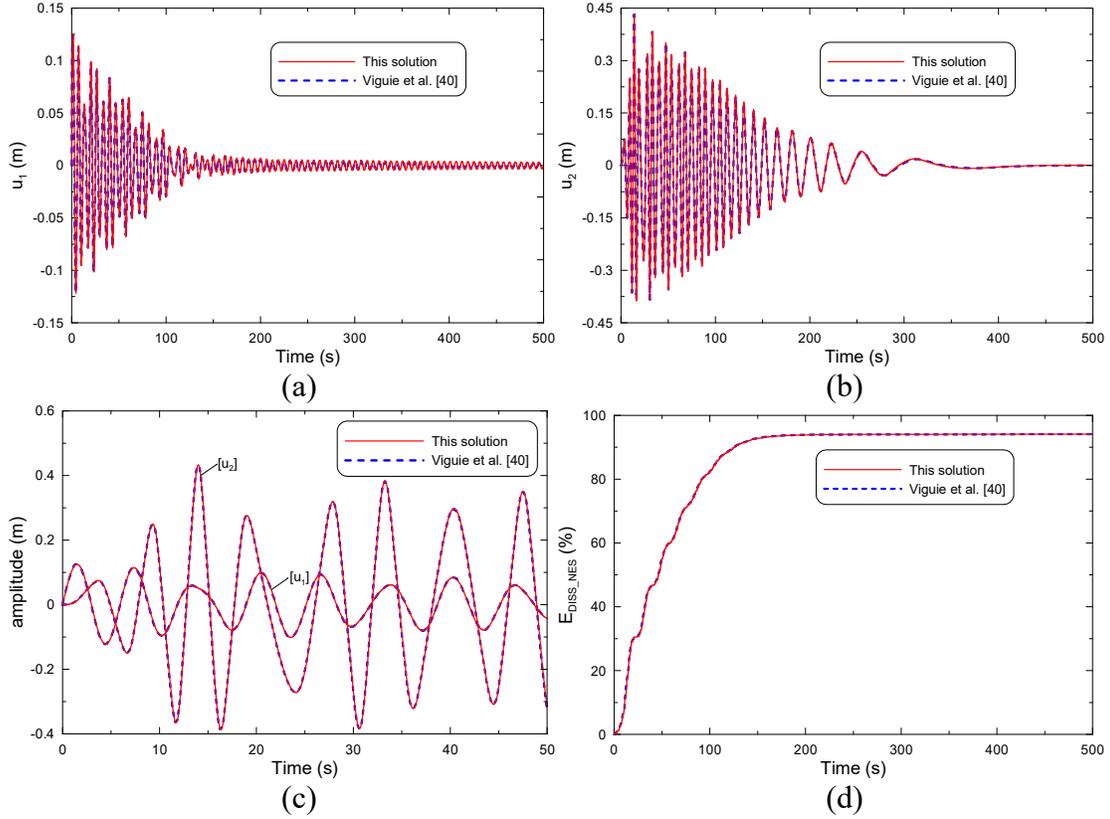

**Fig. 5.** (a) The Duffing oscillator response, (b) NES response; (c) close-up of the Duffing oscillator and NES responses (d) percentage of initial energy dissipated in the NES.

*7.2 HNES performance*

This example is devoted to the investigation of the HNES performance in order to exemplify its salient features. In this work, the energy initially induced into the system by impulse which is represented by the corresponding initial velocity, varies from low ($v_0 = 0.1\,\text{m/s}$) to very high ($v_0 = 2\,\text{m/s}$) energy levels.

The parameters of the primary system are: $m_1 = 1\,\text{kg}$, $k_1 = 1\,\text{N/m}$, and $c_1 = 0.001\,\text{Ns/m}$, whereas the parameters of the HNES are: $m_2 = 0.05\,\text{kg}$, $c_2 = 0.01\,\text{Ns/m}$, $k_{nl2} = 1\,\text{N/m}^3$. The other parameters of the Bouc-Wen model ($a$, $k_2$, $D$, $\gamma$, $n$) are selected through a series of parametric simulations of 500 seconds in order to achieve an optimum performance. In general, the optimal parameters vary and a successful design depends on a specific range of energy values. As is the case with the ordinary NES, in the proposed HNES there exists a threshold of input energy below which the HNES practically does not function. This threshold is approximately 0.1 m/s. Detailed investigation of HNES' behavior below this threshold is beyond the scope of this paper and will be the subject of a future study.

In order now to reduce the free parameters and to ensure a smooth transition from pre-yield to post-yield branch the value of the exponential parameter $n$ was set to 1 ($n=1$) [28]. Remarkable results are achieved concerning the HNES performance which is enhanced when the linear spring stiffness takes on negative values.

First, in order to find the optimum value of the dimensionless parameter $\gamma$ a series of solutions has been performed for $a = -0.3$, $k_2 = 0.1$, $D = 0.1$. Fig. 6 shows contour plot of the energy dissipated in the HNES against the impulse magnitude $v_0$ and the dimensionless parameter $\gamma$ of the Bouc-Wen model. It is obvious that for $\gamma = 0.5$ the HNES performance is within the range $88\% < \eta < 97\%$. In this case, the Bouc-Wen model predicts linear unloading paths [28], [29].

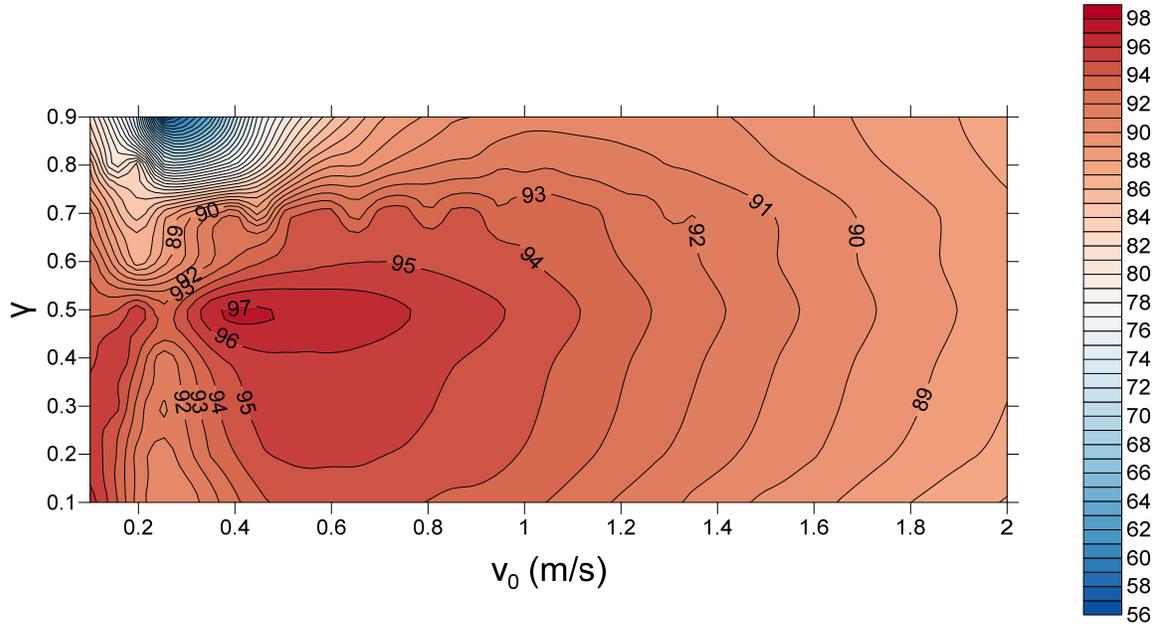

**Fig. 6.** Contour plot of the energy dissipated in the HNES against the impulse magnitude $v_0$ and the dimensionless parameter $\gamma$.

Subsequently, the influence of the parameter $a$ of the Bouc-Wen model is investigated.

For positive values of $a$ the linear spring obtain positive stiffness, whereas for negative values it obtains negative stiffness, leading to a true softening behavior. Fig. 7 depicts the contour plot of the energy dissipated in the HNES against the impulse magnitude $v_0$ and the parameter $a$ of the Bouc-Wen model for $\gamma = 0.5$, $k_2 = 0.1$, $D = 0.1$. Moreover, for clarity in illustration, Fig. 8 shows the percentage of the energy dissipated in the HNES for various values of the parameter $a$. It is clear that the great performance of the HNES is evidenced when $a$ takes on negative values. On the contrary, HNES shows poor performance as the positive values of the parameter $a$ increase, but still achieves a very good percentage of energy dissipation. More specifically, for $a = -0.2$ HNES performance is within the range $88\% < \eta < 97\%$, for $a = -0.1$ HNES performance is within $87\% < \eta < 97\%$, for $a = 0.0$ HNES performance is within $86\% < \eta < 96\%$, and for $a = 0.1$ HNES performance is $84\% < \eta < 96\%$.

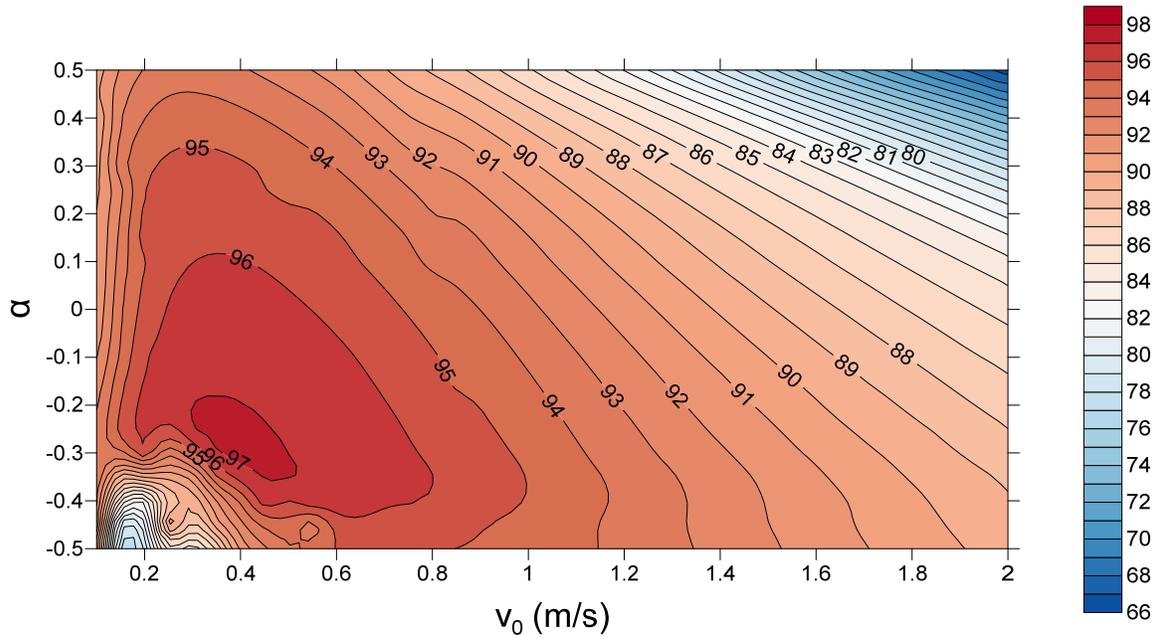

**Fig. 7**. Contour plot of the energy dissipated in the HNES against the impulse magnitude $v_0$ and the parameter $a$.

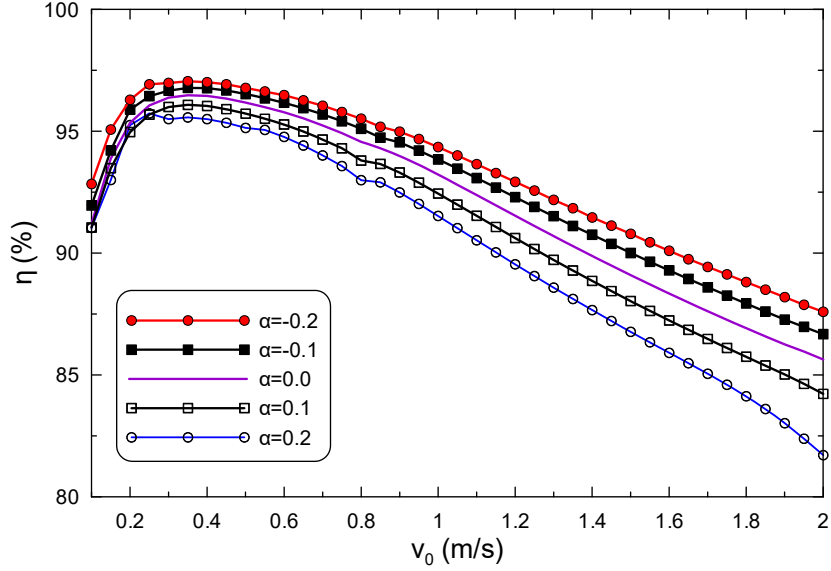

**Fig. 8**. Percentage of the energy dissipated in the HNES for various values of the parameter $a$.

AL-Shudeifat [4] has also studied the shock mitigation of the same primary system employing the Type VII NES with negative linear and nonlinear stiffness elements. Those negative linear and nonlinear stiffness components in the NES were realized through the geometric nonlinearity of the transverse linear springs. He observed that a further enhancement of the NES performance is achieved when the springs are initially loaded. The best design achieved by AL-Shudeifat [4] together with our best design for $a=-0.2$, $k_2 = 0.1$, $D = 0.1$, $\gamma = 0.5$ are presented in Fig. 9. It is clearly seen that Type VII NES performs better for low initially energy level ($0 < v_0 \leq 0.25 \, \text{m/s}$) while HNES has better performance for intermediate energy level ($0.25 < v_0 \leq 0.5 \, \text{m/s}$) initially induced into the system. For initial velocities $0.5 < v_0 \leq 2 \, \text{m/s}$ results from our design are only available which validate once more HNES great performance. It is remarkable that HNES dissipates the 97.05% of the input energy induced into the linear structure when $v_0 = 0.35 \, \text{m/s}$ (intermediate initial energy level) and 87.59% of the input energy induced into the linear structure when $v_0 = 2 \, \text{m/s}$ (very high initial energy level). To verify this observation, the time histories for the primary mass displacement are plotted in Fig. 10 for $v_0 = 0.35 \, \text{m/s}$ and $v_0 = 2.0 \, \text{m/s}$, respectively.

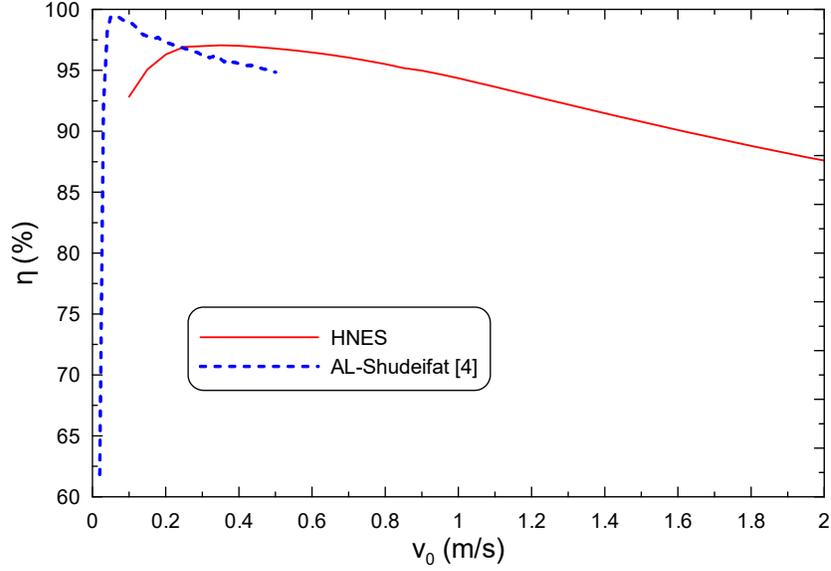

**Fig. 9**. Percentage of the energy dissipated by the HNES as compared to Type VII NES.

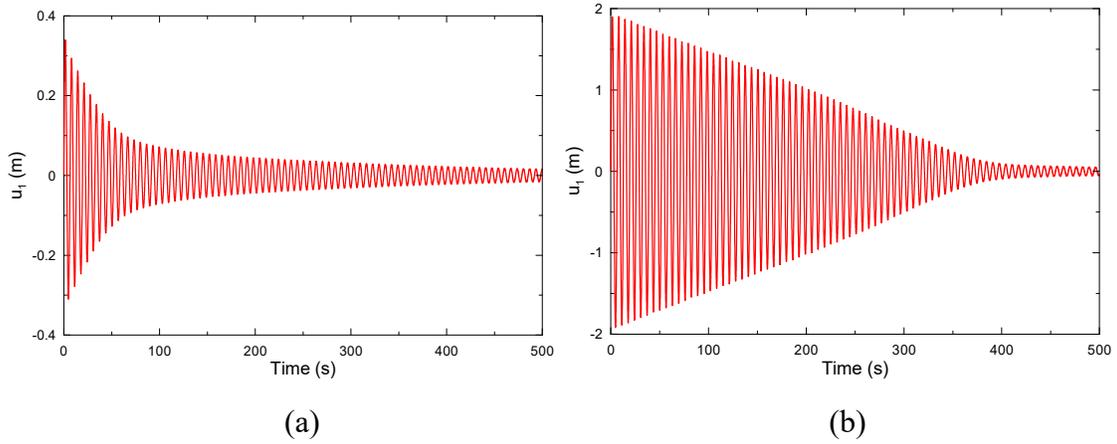

**Fig. 10**. Primary mass displacement for (a) $v_0 = 0.35\,\text{m/s}$ and (b) $v_0 = 2.0\,\text{m/s}$.

## 8. Conclusions

In this paper, the behavior of a new Hysteretic Nonlinear Energy Sink (HNES) coupled to a linear primary oscillator has been investigated in shock mitigation. Apart from a small mass and a nonlinear elastic spring of the Duffing oscillator, the HNES is also comprised of a purely hysteretic and a linear elastic spring of potentially negative stiffness, connected in parallel. The main conclusions that can be drawn from this investigation are:

- HNES is a robust Nonlinear Energy Sink which can efficiently dissipate a large fraction of the energy initially induced into the system.
- Remarkable results are achieved concerning the HNES performance which is enhanced when the linear spring stiffness takes on negative values and is validated for a wide range of initial input energies.
- On the contrary, HNES shows poor performance as the positive values of the parameter $a$ increase, but still achieves a very good percentage of energy dissipation.
- Type VII NES performs better for low initially energy level ($0 < v_0 \leq 0.25\,\text{m/s}$) while HNES has better performance for intermediate energy level ($0.25 < v_0 \leq 0.5\,\text{m/s}$) initially induced into the system.
- It is remarkable that HNES dissipates the 97.05% of the input energy induced into the linear structure when $v_0 = 0.35\,\text{m/s}$ (intermediate initial energy level) and 87.59% of the input energy induced into the linear structure when $v_0 = 2\,\text{m/s}$ (very high initial energy level).
- Finally, the salient feature of the proposed device great performance lies in the dominant role of the hysteresis which in addition to the negative stiffness element renders HNES an alternative, yet superior NES.

## 9. References

[1] Watts P. On a method of reducing the rolling of ships at sea. Transactions of the Institution of Naval Architects 1883;24:165–190.
[2] Frahm H. Device for Damping Vibrations of Bodies. US patent #989958, 1909.
[3] Weber B, Feltrin G. Assessment of long-term behavior of tuned mass dampers by system identification. Eng Struct 2010;32:3670–3682.
[4] AL-Shudeifat M. Highly efficient nonlinear energy sink. Nonlinear Dyn 2014;76:1905–1920.
[5] Boroson E, Missoum S, Mattei P-O, Vergez C. Optimization under uncertainty of parallel nonlinear energy sinks. J Sound Vibr 2017;394:451–464.